\pdfoutput=1
\documentclass[twocolumn,english,aps,pra,longbibliography,superscriptaddress,floatfix]{revtex4-1}

\usepackage[utf8]{inputenc}
\usepackage[T1]{fontenc}
\usepackage{amsmath,amssymb,amsfonts,amsthm}
\usepackage{graphicx}
\usepackage{booktabs}
\usepackage{soul}
\usepackage[activate={true,nocompatibility},tracking=true,kerning=true,spacing=true,factor=1100]{microtype}
\SetTracking{encoding={*}, shape={*}, series={*}, size={*}}{-10}
\usepackage[dvipsnames]{xcolor}
\usepackage[ruled,vlined]{algorithm2e}
\usepackage[colorlinks=true,urlcolor=blue,citecolor=blue,linkcolor=blue]{hyperref}

\begin{document}

\title{Scaling Quantum Optimization to the Thousand-Qubit Scale with Distributed Quantum Sampling}

\author{Amana Liaqat}
\affiliation{Qoro Quantum Ltd., United Kingdom}

\author{Ahmed Darwish}
\affiliation{Qoro Quantum GmbH, Germany}

\author{Stephen DiAdamo}
\email{stephen@qoroquantum.de}
\affiliation{Qoro Quantum GmbH, Germany}

\author{Dan Holme}
\affiliation{Qoro Quantum Ltd., United Kingdom}

\author{Kieran McDowall}
\affiliation{National Quantum Computing Centre, United Kingdom}
\affiliation{Quantum Software Lab, The University of Edinburgh, United Kingdom}

\author{Emre Sahin}
\affiliation{Hartree Centre, STFC, United Kingdom}

\author{Naeimeh Mohseni}
\affiliation{E.ON Digital Technology GmbH, Germany}

\author{Giorgio Cortiana}
\affiliation{E.ON Digital Technology GmbH, Germany}

\author{Corey O'Meara}
\email{corey.omeara@eon.com}
\affiliation{E.ON Digital Technology GmbH, Germany}

\begin{abstract}
The Minimum Birkhoff Decomposition (MBD) seeks a sparse weighted sum of matchings and is a challenging optimization problem with applications in network scheduling and energy trading. We scale a quantum-assisted decomposition method by combining single-layer QAOA sampling with Extended Fully-Corrective Frank-Wolfe (E-FCFW) optimization, spectral graph partitioning, and greedy feasibility repair. We demonstrate the pipeline on the 1,354-node PEGASE bus test case (representing a 1,354-bus transmission grid with 1,710 transmission lines), whose 1,710 edges define a native 1,710-variable matching optimization problem requiring 1,710 qubits in the unpartitioned encoding. Partitioning enables distributed execution on IBM superconducting quantum processors. The best reported hardware result uses a 50-qubit partition bound, while reducing partitions to 20 qubits degrades convergence in the partition-size comparison. Larger partitions are more demanding for matrix product state (MPS) simulation, and bond-dimension tests show that truncating quantum correlations reduces candidate quality. The results therefore motivate retaining a substantive quantum sampling task within each partition as the overall problem scales. On the PEGASE-1354 benchmark, repaired QAOA samples achieve lower residual decomposition error than both simulated annealing and uniform random sampling baselines, while spatial circuit packing reduces hardware execution time by approximately a factor of three. These results demonstrate that combining spectral partitioning, spatial circuit packing, and classical feasibility repair provides an executable path for deploying gate-based quantum sampling on thousand-variable constrained optimization problems on current hardware.

\end{abstract}

\maketitle

\section{Introduction}
\label{sec:intro}

Network scheduling, optical switching, and peer-to-peer energy trading \cite{mohseni2026evidence, bucher2024evaluating} require weighted demands to be served through compatible sets of connections \cite{blenninger2024q,ganeshamurthy2024next}. Representing each set as a matching gives a graph decomposition problem: reconstruct the demand graph with as few weighted matchings as possible. Fewer matchings reduce switching and reconfiguration overhead. This extends the Birkhoff--von Neumann decomposition from bipartite graphs to general network topologies \cite{brualdi1982notes}.

Minimum Birkhoff Decomposition (MBD) is NP-hard~\cite{dufosse2016notes} and can challenge classical optimization at small instance sizes. Its inclusion in the Quantum Optimization Benchmarking Library (QOBLIB) identifies it as a relevant test problem for progress toward quantum advantage in optimization \cite{koch2026quantum}. The difficulty lies in selecting a sparse decomposition; finding an individual maximum-weight matching remains polynomial-time solvable. Quantum sampling is therefore considered here as a way to enrich the candidate pool of a classical decomposition algorithm.

Building on Extended Fully-Corrective Frank-Wolfe (E-FCFW) \cite{pennington2025boosting}, we use single-layer ($p=1$) QAOA to supply diverse matching candidates and optimize their weights classically. Fixed, offline-calibrated angles avoid variational tuning during execution. Although complexity-theoretic results motivate studying shallow QAOA sampling \cite{farhi2016quantum}, they do not establish an advantage for the instances studied here.

\begin{figure*}[ht]
\centering
\resizebox{\textwidth}{!}{\includegraphics{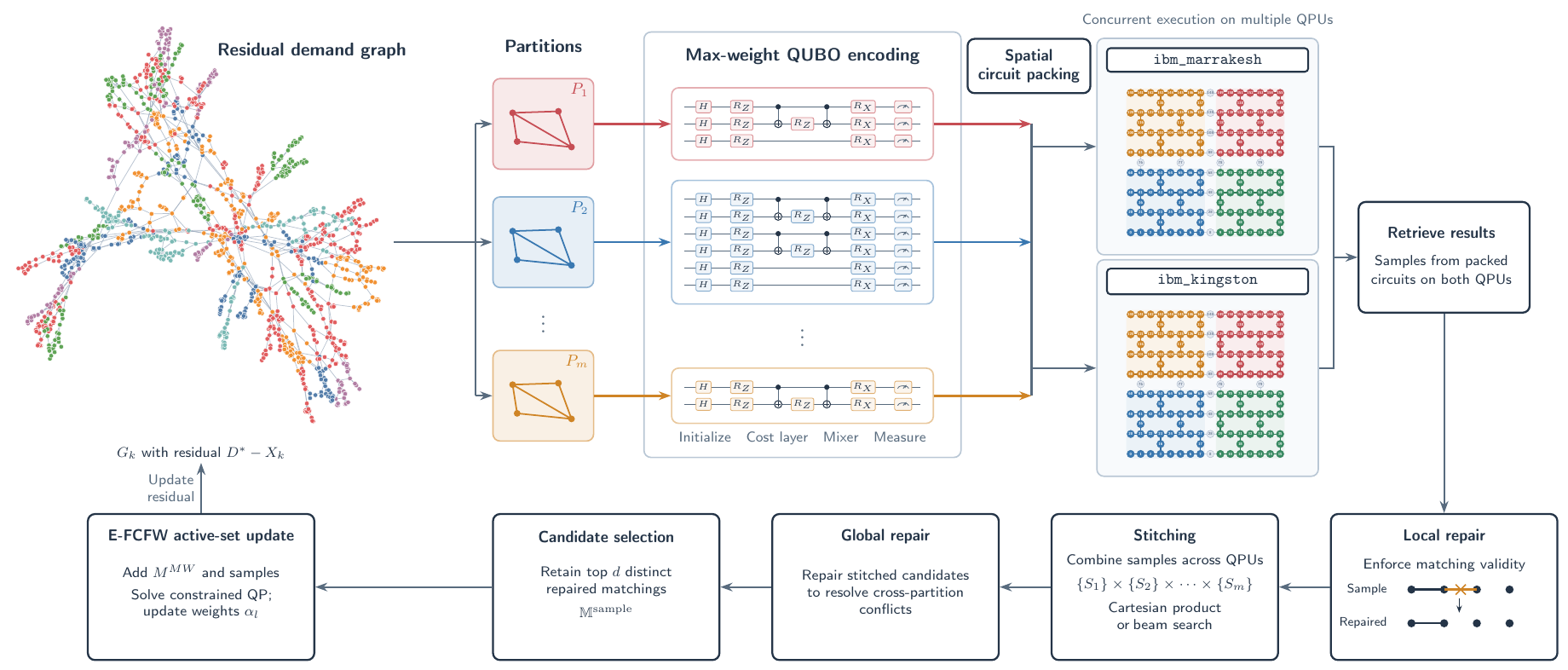}}
\caption{Residual graphs are partitioned into QAOA subproblems and dispatched across IBM QPUs using spatial circuit packing. Samples are stitched, repaired to enforce matching feasibility, and added to the classical E-FCFW active set.}
\label{fig:partitioning_workflow}
\end{figure*}

While divide-and-conquer and multilevel methods have scaled unconstrained QUBO subproblems on quantum processors to tens of thousands of variables \cite{maciejewski2024multilevel,li2022large}, our contribution is to scale quantum-assisted constrained optimization for graph decomposition beyond the width of an individual processor. Spectral partitioning reduces circuit width and routing overhead; stitching and greedy repair reconstruct feasible global candidates; and spatial circuit packing executes independent subproblems concurrently. Figure~\ref{fig:partitioning_workflow} summarizes the workflow. We examine its dependence on matrix product state (MPS) bond dimension and partition size, then demonstrate hardware execution on the 1,354-node, 1,710-edge PEGASE bus test case \cite{thurner2018pandapower}, which models a realistic 1,354-bus, 1,710-line European high-voltage transmission network developed by the Pan-European Grid Advanced Simulation and State Estimation (PEGASE) project. The central question is whether partitioning can make execution practical while retaining useful quantum sampling within each partition.

\section{Problem Formulation}
\label{sec:problem}

In this section, we formulate the problem of decomposing general network demand graphs into sparse linear combinations of valid matchings. We first define the MBD as a cardinality-minimization problem over symmetric doubly substochastic matrices. We then review the Extended Fully-Corrective Frank--Wolfe (E-FCFW) optimization framework, which augments classical Frank--Wolfe steps with diverse candidate sampling to achieve sparser graph representations.

\subsection{Minimum Birkhoff Decomposition}
\label{subsec:mbd}

Let $G(V,E)$ denote a weighted, undirected graph with edge weights $w(e) \ge 0$ representing demand or connection intensity. A subgraph $M \subseteq E$ is a valid matching if no two edges in $M$ share a common vertex. We represent $G$ as a symmetric doubly substochastic matrix $D^* \in [0,1]^{n \times n}$ with non-negative entries satisfying:
\begin{equation}
\sum_{j=1}^{n} D^*_{ij} \le 1, \quad \sum_{i=1}^{n} D^*_{ij} \le 1 \quad \forall i,j,
\end{equation}
with zero diagonal entries ($D^*_{ii} = 0$, no self-loops) and symmetric weights ($D^*_{ij} = D^*_{ji}$).

A matching matrix $P$ belongs to the feasible set $\mathcal{P}$:
\begin{equation}
\mathcal{P} := \{U \in \{0,1\}^{n \times n} \mid U = U^T, \, U \mathbf{1} \preceq \mathbf{1}, \, U_{ii} = 0 \,\, \forall i\},
\end{equation}
where $\mathbf{1}$ denotes the vector of all ones. The MBD problem seeks to minimize the number of constituent matchings needed to reconstruct $D^*$, formulated as an $\ell_0$-norm minimization:
\begin{align}
\min_{\alpha, P_l} \quad & \|\alpha\|_0 \label{eq:mbd_obj}\\
\text{s.t.} \quad & D^* = \sum_{l=1}^{m} \alpha_l P_l, \quad P_l \in \mathcal{P}, \nonumber \\
& \alpha_l \ge 0, \quad \sum_{l=1}^{m} \alpha_l \le 1, \nonumber
\end{align}
where $\|\alpha\|_0 = |\{l : \alpha_l > 0\}|$ counts the non-zero weights across a decomposition of length $m$.
An exact decomposition with $\sum_l\alpha_l\le1$ requires $D^*$ to lie in the convex hull of matching matrices (including the empty matching). On non-bipartite graphs, double substochasticity alone is insufficient; otherwise, the framework below computes a sparse approximation. Even in the classical doubly stochastic setting, minimizing decomposition cardinality is NP-hard \cite{dufosse2016notes}. The hard MBD instances documented in QOBLIB motivate sampling-based approaches beyond greedy decomposition \cite{koch2026quantum,durkovic2017birkhoff}.

\subsection{The Extended Fully-Corrective Frank-Wolfe Framework}
\label{subsec:efcfw}

We seek sparse approximations using the Frank-Wolfe (FW) framework \cite{combettes2023revisiting, valls2021birkhoff}. For graph decomposition, the objective function is the scaled squared Frobenius norm distance between the target demand $D^*$ and the approximation matrix $D$:
\begin{equation}
f(D) := \frac{1}{n^2} \|D - D^*\|_F^2.
\end{equation}
At iteration $k$, standard FW linearizes $f(D)$ around the current iterate $D_k$ and queries a Linear Minimization Oracle (LMO):
\begin{equation}
M^{MW} \in \arg\min_{U \in \mathcal{P}} \sum_{i,j=1}^{n} \nabla f(D_k)_{ij} U_{ij},
\end{equation}
which maps to finding a maximum-weight matching on the residual graph $D^* - D_k$ \cite{edmonds1965paths}.

Fully-Corrective Frank-Wolfe (FCFW) re-optimizes weights over all previously collected matchings in an active set $\mathbb{M}$. Defining the substochastic unit simplex $\Delta_m := \{u \in \mathbb{R}_{\ge0}^m : \sum_{i=1}^m u_i \le 1\}$, the weights are updated via:
\begin{equation}
(\alpha_1, \dots, \alpha_{|\mathbb{M}|}) = \arg\min_{u \in \Delta_{|\mathbb{M}|}} \frac{1}{n^2} \left\| D^* - \sum_{i=1}^{|\mathbb{M}|} u_i M_i \right\|_F^2.
\end{equation}

Deterministic oracle calls can yield similar matchings, motivating additional candidate diversity. The E-FCFW algorithm \cite{pennington2025boosting} addresses this by augmenting the candidate pool at each iteration: alongside the deterministic maximum-weight matching $M^{MW}$ (which supplies a Frank-Wolfe descent direction before cardinality restriction), a sampling subroutine generates $d$ diverse candidate matchings $\mathbb{M}^{\text{sample}}$, updating the active pool as $\mathbb{M} \leftarrow \mathbb{M} \cup \{M^{MW}\} \cup \mathbb{M}^{\text{sample}}$. E-FCFW then enforces an explicit $\ell_0$-norm constraint:
\begin{equation}
\min_{u \in \Delta_{|\mathbb{M}|}} \frac{1}{n^2} \left\| D^* - \sum_{i=1}^{|\mathbb{M}|} u_i M_i \right\|_F^2 \quad \text{s.t.} \quad \|u\|_0 \le k+1.
\end{equation}
The cardinality restriction retains at most $k+1$ positive weights. Weight optimization on a fixed support is a convex quadratic program; support selection remains combinatorial.

\section{Methodology}
\label{sec:methods}

Each E-FCFW iteration partitions the residual graph, samples local QAOA circuits, reconstructs global matchings, and updates the decomposition weights. We describe the essential components here; calibration, reconstruction settings, pseudocode, and scheduling details are collected in Appendix~\ref{app:methods}.

\subsection{QAOA Matching Sampler}
\label{subsec:qaoa}

Each edge $j\in E$ is encoded by a qubit, with $x_j=1$ indicating its selection. For residual edge weights $w_j$, we use the QUBO
\begin{equation}
C(x)=-\sum_{j\in E}w_jx_j+\lambda\sum_{\substack{j\in E,\\k\in\Gamma(j)}}x_jx_k,
\end{equation}
where $\Gamma(j)$ contains edges sharing an endpoint with $j$. The sum follows the ordered-pair convention, counting each conflicting pair twice. The penalty $\lambda=0.2$ discourages collisions; greedy repair subsequently enforces feasibility. Substituting $x_j=(\mathbb{I}-\hat Z_j)/2$ defines the diagonal cost Hamiltonian $\hat H_C$ and the single-layer state
\begin{equation}
|\psi(\beta,\gamma)\rangle= e^{-i\beta\sum_{j=1}^{N}\hat X_j}e^{-i\gamma\hat H_C}|+\rangle^{\otimes N},
\end{equation}
with $N=|E|$. The sampler supplies a pool of candidates rather than a single optimal matching.

Angles are calibrated offline and transferred across related subgraphs, following prior work on QAOA parameter transfer \cite{pelofske2024scaling,pennington2025boosting}. We use $(\gamma_0,\beta_0)=(0.5,-0.5)$ for the 120-node benchmark and approximately $(2.49, 0.16)$ for the PEGASE bus test case. They remain fixed throughout execution, eliminating online parameter optimization.

\subsection{Partitioning and Reconstruction}
\label{subsec:partitioning}

While recursive Kernighan--Lin bisection was evaluated in preliminary benchmarking, multi-way spectral partitioning produced subgraphs with fewer cut edges and superior downstream convergence, and was therefore adopted throughout this work. Multi-way spectral partitioning clusters affinity-matrix eigenvectors using $k$-means and recursively refines oversized clusters. Throughout, $p_s$ denotes the maximum number of edges per partition, hence the maximum circuit qubit count, rather than a bound on graph vertices. Spectral partitioning groups strongly connected regions while limiting the number of connections removed from local Hamiltonians.

The residual graph is repartitioned at each iteration. Cut edges cannot be sampled by that iteration's local circuits, but their residual weights can influence later partitions. This provides a route to recovering omitted connections, without guaranteeing that every edge becomes internal within a finite iteration budget. The global classical maximum-weight matching oracle also remains available to the outer solver.

Local samples are combined by batched Cartesian products or pruned beam search. We tested both methods and used batched Cartesian products, as it maintains sample validity more effectively. Recombination can introduce conflicts at cut vertices shared across partition boundaries, alongside any intra-partition collisions permitted by the soft QUBO penalty. 

To restore feasibility without introducing unselected edges, greedy repair sorts the candidate edge set $\tilde E=\{e_j:x_j=1\}$ by descending residual weight $w(e_j)=(D^*-D_k)_{uv}$ and accepts an edge if and only if both endpoints are currently unoccupied. For $|\tilde E|$ selected edges, this pruning requires $O(|\tilde E|\log|\tilde E|)$ time and guarantees matching feasibility. Because repair alters the sample distribution without adding edges beyond those proposed by the quantum sampler, the experiments compare QAOA and random samples under the identical repair procedure to isolate the contribution of sampling.

\subsection{Classical Update and Hardware Execution}

The best distinct repaired candidates are added to the active pool together with a global maximum-weight matching. E-FCFW updates the nonnegative weights subject to its cardinality bound and removes zero-weight matchings. The loop stops at the target error, a relative-improvement plateau, or the iteration limit. Decomposition length counts retained positive-weight matchings and need not equal the iteration index. Algorithm~\ref{alg:partitioned_qaoa} gives the full workflow.

Qoro's Divi framework \cite{divi2024} coordinates circuit generation, sampling, reconstruction, and execution on \texttt{ibm\_kingston} and \texttt{ibm\_marrakesh}. Qoro's scheduler packs independent circuits into disjoint regions of the 156-qubit Heron lattice, using otherwise idle capacity and buffer qubits between subproblem boundaries to reduce dispatch rounds (Fig.~\ref{fig:partitioning_workflow}).

\section{Experimental Results}
\label{sec:results}

In this section, we evaluate the performance of the distributed QAOA sampling framework on graph decomposition benchmarks. We compare partitioned E-FCFW across varying partition bounds against classical simulated annealing and uniform random sampling baselines, implemented both via MPS tensor network simulation and directly on superconducting quantum processors. We further isolate the impact of classical feasibility repair on solution quality and demonstrate throughput speedups achieved through multi-circuit spatial packing on an IBM Heron device.

\subsection{Experimental Setup}

We evaluate a synthetic Heron-topology graph with 120 vertices and 132 edges, followed by the PEGASE-1354 bus test case with 1,354 vertices and 1,710 edges. The generation of the graphs is detailed in \cite{pennington2025boosting}. QAOA circuits use $p=1$ and 25,000 shots per partition per round. Simulated annealing (SA) \cite{kirkpatrick1983optimization} uses 1,500 reads per partition, each with 1,000 sweeps. Uniform random samples pass through the same stitching and repair as QAOA and SA. These are specified sampling budgets, not matched computational costs. 

Hardware execution relied on heuristic transpilation passes targeting minimal two-qubit gate depth. While active error mitigation techniques—specifically dynamical decoupling using the $X_pX_m$ pulse sequence and gate twirling with 32 randomizations—were evaluated, empirical testing showed they degraded candidate quality on these partitioned workloads. Consequently, hardware runs omitted pulse-level mitigation in favor of depth-minimized compilation coupled with downstream greedy repair (Appendix~\ref{app:calibration}).

All samplers supply the top five distinct valid candidates to E-FCFW. The stopping criteria are normalized error $\epsilon=10^{-6}$ or relative cost improvement at most $10^{-3}$. Standalone greedy Birkhoff decomposition (BKF) is also included as a baseline. Appendix~\ref{app:calibration} gives the calibration and stopping details.

\subsection{Classical Simulation and Hardware Convergence}
\label{subsec:exp_mps}

To classically emulate and diagnose the distributed quantum sampling pipeline, we simulate the QAOA circuits using MPS tensor networks via Maestro \cite{bertomeu2025maestro}. In this representation, the MPS bond dimension $\chi$ bounds the bipartite entanglement entropy across any cut by $\log\chi$, directly controlling the amount of quantum correlation retained in the simulated state. This allows us to systematically probe how truncating entanglement impacts candidate sample quality and downstream decomposition performance. 

\begin{figure*}[ht]
\centering
\resizebox{\textwidth}{!}{\includegraphics{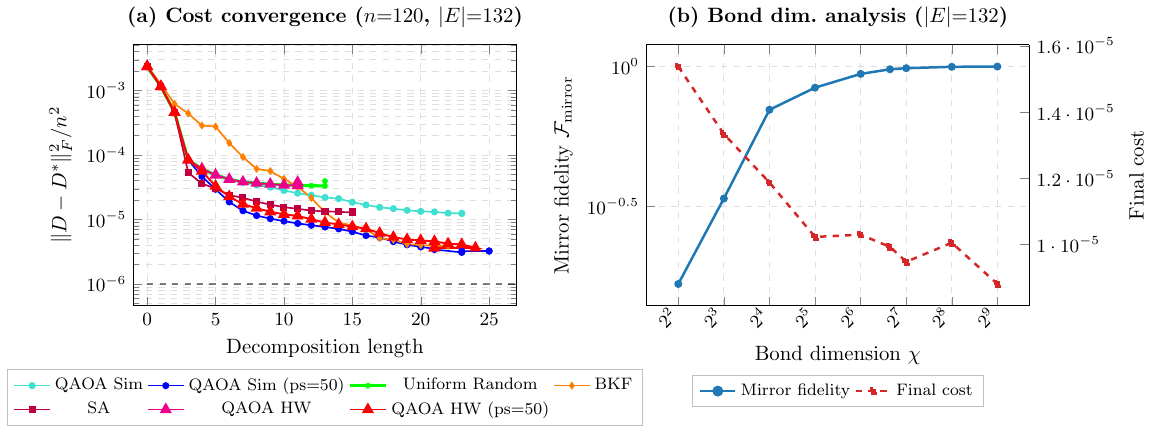}}
\caption{Benchmark on synthetic demand graphs inspired by heavy-hex topology ($n=120$, $|E|=132$). (a) Normalized approximation cost versus decomposition length for hardware QAOA, MPS simulation, and classical baselines. All methods except greedy Birkhoff (BKF) supply candidates to E-FCFW. (b) Mirror fidelity and final cost versus MPS bond dimension $\chi$: retaining more entanglement improves simulation fidelity and candidate quality.}
\label{fig:cost-fidelity}
\end{figure*}

We compare final decomposition cost with mirror fidelity: the probability of recovering the all-zero state after simulating a state-preparation circuit and its inverse at finite $\chi$ (Appendix~\ref{app:mirror}). Higher bond dimensions improve both mirror fidelity and final cost in Fig.~\ref{fig:cost-fidelity}(b). Thus, retaining the correlations lost at low $\chi$ matters for this sampling task. This diagnostic measures the limitations of the tested MPS approximation; it does not establish hardness for all classical simulation methods.

Figure~\ref{fig:cost-fidelity}(a) compares hardware and simulated QAOA with the classical baselines. Hardware QAOA tracks the simulated convergence and achieves lower costs than random sampling and simulated annealing. Partitioning at $p_s=50$ retains convergence close to the unpartitioned hardware run, supporting its use to reduce circuit size without losing the useful sampling behavior in this instance. SA makes initial progress but triggers the plateau criterion at round 18, with decomposition length 15; QAOA continues to improve. These observations compare the tested samplers and budgets, rather than establishing a general separation from classical optimization.

\subsection{Partition Size and Circuit Density}
\label{subsec:exp_part}

The bounds $p_s\in\{20,50,80\}$ expose a trade-off between graph fragmentation and circuit complexity. Aggressive partitioning ($p_s=20$) yields uniformly shallow circuits that are readily approximated at low MPS bond dimension ($\chi\le16$), but convergence deteriorates as more edges are cut. Relaxing the bound to $p_s=50$ and $p_s=80$ preserves better convergence while reducing hardware requirements relative to the unpartitioned problem, but circuit depth and routing overhead become heavily dependent on local subgraph topology rather than qubit count alone.

Table~\ref{tab:hardware_depths} makes this topological dependence explicit. An 80-qubit circuit has transpiled two-qubit depth 67, whereas circuits with only 42--52 qubits reach depths 110--169. Because qubits represent edges rather than vertices, two qubits interact whenever their edges share an endpoint vertex. While the synthetic graph is generated on Heron's degree-3 lattice, its line graph (the interaction graph of the edge-qubits) introduces triangles and higher local connectivity. Subgraphs that contain dense loop structures or clustered degree-3 junctions produce tightly coupled line graphs that require extensive SWAP routing when mapped back onto Heron's physical qubits.

Conversely, partitions that form predominantly linear paths or open trees route with minimal SWAP overhead. Consequently, logical two-qubit depth tracks local interaction density rather than qubit count, and directly governs the transpiled routing depth. Recent experimental demonstrations on IBM quantum computers reach two-qubit gate depths of up to 452 with reliable output at this qubit count \cite{hartnett2026fast}. However, qubit count alone is an inadequate partitioning objective: local interaction density and hardware embedding determine execution cost. The next scaling step should control partition density alongside size, retaining enough interactions to challenge classical sampling while keeping routed depth manageable.

\begin{table}[ht]
\caption{Circuit width and two-qubit depth before (logical) and after transpilation to IBM Heron. The partition bound $p_s$ counts edges (qubits). Logical two-qubit depth reflects the maximum degree of interacting edge pairs, which governs transpiled depth on heavy-hex hardware.}
\label{tab:hardware_depths}
\begin{ruledtabular}
\begin{tabular}{crrr}
$p_s$ & Qubits & 2Q depth & Transpiled 2Q depth \\
\colrule
80 & 80 & 28 & 67 \\
   & 52 & 66 & 169 \\
   & 46 & 62 & 133 \\
\colrule
50 & 49 & 74 & 164 \\
   & 42 & 50 & 110 \\
   & 50 & 38 & 119 \\
\colrule
20 & 20 & 14 & 24 \\
\end{tabular}
\end{ruledtabular}
\end{table}

\subsection{Deployment on PEGASE-1354}
\label{subsec:exp_utility}

The PEGASE-1354 transmission graph would require 1,710 qubits under an unpartitioned edge encoding. Spectral partitioning at $p_s=50$ instead produces 41 subgraphs for distributed execution (illustrated conceptually in Fig.~\ref{fig:partitioning_workflow} and detailed in the complete 41-subgraph layout of Fig.~\ref{fig:pegase1354_partitions}, Appendix~\ref{app:deployment}). Spatial packing combines independent subproblems concurrently on a processor; specifically, two 38-qubit and two 34-qubit circuits occupy 144 of the 156 available qubits on \texttt{ibm\_marrakesh}, separated by idle buffer qubits (mapped to physical hardware in Fig.~\ref{fig:heron_circuit_packing}).

\begin{figure*}[ht]
\centering
\includegraphics[width=\textwidth]{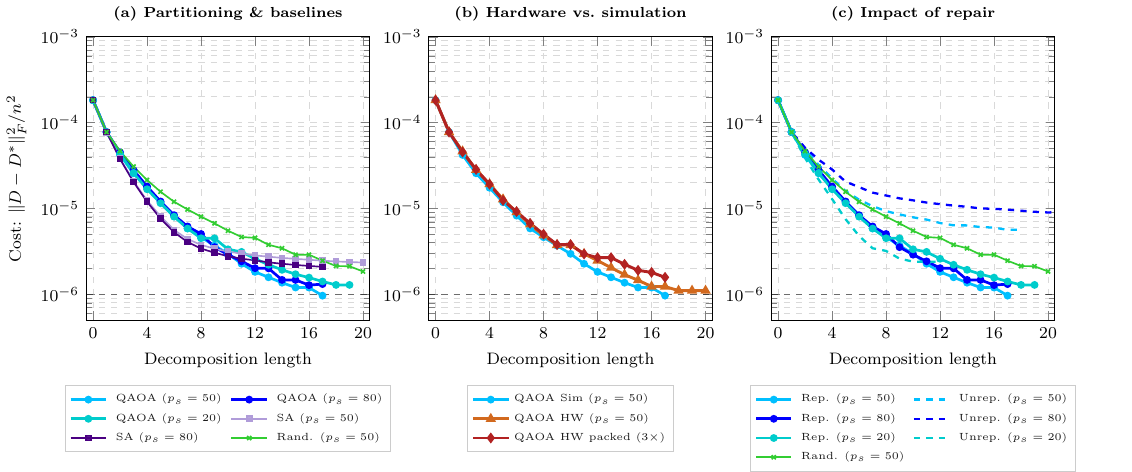}
\caption{Cost convergence of partitioned E-FCFW versus decomposition length on simulation and quantum hardware compared with classical baselines, with and without repair. (a) Partitioning bounds $p_s \in \{20, 50, 80\}$ evaluated via MPS simulation ($\chi=128$), compared with partitioned simulated annealing (SA) and uniform random sampling baselines (all using feasibility repair). (b) Hardware execution on IBM Heron (\texttt{ibm\_marrakesh}) tracking 128-bond-dimension MPS simulation, with spatial circuit packing yielding a $3\times$ execution speedup with minimal degradation in candidate quality. (c) Convergence with and without greedy feasibility repair; discarding invalid matchings without repair stalls QAOA convergence above $10^{-5}$, while uniform random sampling without repair yields zero valid matchings. Random sampling is shown with repair. Cost is $\|D-D^*\|_F^2/n^2$; decomposition length counts retained matchings, not outer iterations.}
\label{fig:hardware-repair-run}
\end{figure*}

Figure~\ref{fig:hardware-repair-run}(a) compares convergence under partition bounds $p_s \in \{20, 50, 80\}$ using MPS tensor network simulation with bond dimension set to $\chi=128$, compared with partitioned SA and repaired uniform random sampling baselines. Partitioning at $p_s=50$ and $p_s=80$ significantly improves on $p_s=20$, which cuts too many edges and stalls early. Under identical $p_s=50$ partitioning, QAOA sampling achieves lower residual decomposition error than simulated annealing and random sampling baselines.

Figure~\ref{fig:hardware-repair-run}(b) evaluates hardware execution and spatial packing on \texttt{ibm\_marrakesh}. Unpacked hardware closely tracks 128-bond-dimension MPS simulation, reaching $1.11\times10^{-6}$ at the 20-round limit. Packed execution stops at its round-17 plateau with error $1.58\times10^{-6}$, while reducing measured execution time by approximately $3\times$. The modest difference in final convergence ($4.7\times10^{-7}$) reflects a practical trade-off: spatial packing constrains each circuit to a fixed sub-region, reducing transpiler choice for high-fidelity qubits and exposing circuits to potential spectator crosstalk, in exchange for a three-fold speedup.

Figure~\ref{fig:hardware-repair-run}(c) isolates the impact of feasibility repair. Discarding invalid matchings without repair stalls convergence near $10^{-5}$, whereas random sampling without repair produces zero valid matchings and fails to make any progress. Random samples with repair improve feasibility but retain residual error above $2\times10^{-6}$, whereas QAOA with the same repair reaches $1.1\times10^{-6}$, demonstrating that QAOA provides structured candidates that repair enhances rather than replaces.

\section{Discussion and Conclusion}
\label{sec:discussion}

Partitioned QAOA sampling extends E-FCFW decomposition to a 1,710-edge grid using circuits that fit current quantum hardware. On the PEGASE-1354 bus test case, QAOA with repair improves on random sampling with the same repair, and spatial packing reduces execution time by approximately a factor of three. These results establish an executable distributed workflow for a problem class whose classical difficulty makes it relevant to the search for quantum advantage.

For that goal, the partitions must retain a useful quantum sampling task. Reducing them to small, easily simulated fragments would remove the reason to use a quantum processor. The $p_s=50$ results provide evidence for this requirement: partitioned hardware retains the unpartitioned convergence behavior, while the bond-dimension study shows that aggressive truncation of quantum correlations degrades candidate quality. This is evidence of a useful regime, not proof of classical intractability or quantum advantage.

Beyond partition width, the underlying topology determines where quantum sampling provides distinct utility. While extremely sparse graphs admit efficient classical solutions and dense topologies incur prohibitive SWAP-routing overhead, intermediate-density modular networks—such as power transmission grids—represent the primary target regime. In these modular architectures, classical heuristics frequently become trapped in frustrated local minima across partition cut boundaries, whereas shallow QAOA states maintain the candidate diversity necessary to sustain Frank-Wolfe convergence.

The next step is to target partitions of at least 100 qubits and vary their density systematically: enough connectivity to preserve useful quantum correlations, with transpiled two-qubit depth low enough for reliable execution. Table~\ref{tab:hardware_depths} shows why width and density must be optimized together. 

Preliminary work towards this has been carried out on three benchmark instances from QOBLIB's MBD track up to 4,096 variables~\cite{koch2026quantum}, using partitioned QAOA executed on IBM Heron hardware. The initial results show that the method consistently reaches the target tolerance within 15 outer iterations and achieves substantially sparser decompositions (retaining 50 to 75 matchings) than classical greedy baselines such as Birkhoff+ (which required over 250 matchings). Future work will extend this evaluation across the full benchmark suite with comprehensive runtime accounting and comparisons against strong classical solvers. The objective is to preserve a useful quantum contribution within each partition as the overall problem grows.

\section*{Data and Code Availability}

The pipeline implementation, example Divi workflows, demonstration notebooks, and benchmark datasets will be made available. Code requests may be addressed to the authors.

\section*{Acknowledgements}
We acknowledge support from the UK STFC National Quantum Computing Centre SparQ programme through grant SQPOC02 to Qoro Quantum Ltd., entitled ``Quantum-Enhanced Optimisation for Peer-to-Peer Energy Trading''.

\appendix
\makeatletter
\def\p@subsection{\thesection.}
\makeatother
\section{Extended Methods and Hardware Layouts}
\label{app:methods}

\subsection{Parameter Calibration and Sampling Budgets}
\label{app:calibration}

Offline training uses small prototypes ($N\le16$) with similar topology, degree distribution, and edge-weight encoding to the target family. The prototype calibration requires approximately 100 parallel circuits and about ten minutes of MPS simulation in the reported setup. The resulting angles are transferred to larger circuits instead of being optimized at each E-FCFW iteration.

\begin{figure}[ht]
\centering
\includegraphics[width=\columnwidth]{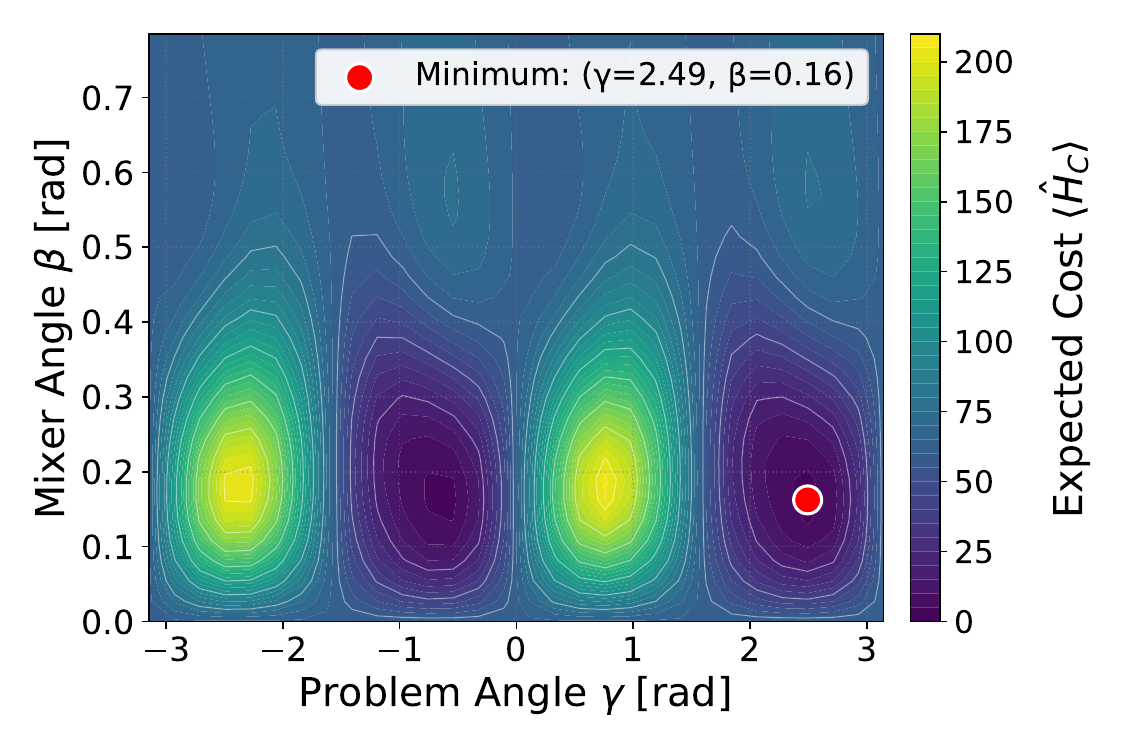}
\caption{QAOA energy landscape for a representative 49-qubit PEGASE partition, simulated with Maestro at bond dimension 128. The marked minimum near $(\gamma_0, \beta_0) = (2.49, 0.16)$ supplies fixed angles for the PEGASE-1354 transmission graph experiments.} 
\label{fig:beta_gamma_heatmap}
\end{figure}

For the PEGASE bus test case, the calibration landscape in Fig.~\ref{fig:beta_gamma_heatmap} is evaluated on a representative 49-qubit partition with Maestro at bond dimension 128, scanning $\gamma \in [-\pi, \pi]$ and $\beta \in [0, \pi/4]$. Its marked minimum near $(\gamma_0, \beta_0) = (2.49, 0.16)$ supplies fixed execution angles [with an equivalent symmetric basin near $(-0.65, 0.16)$ due to $\pi$-periodicity in $\gamma$]. Finite bond dimension makes this an approximate MPS calibration; bond dimension 128 alone does not guarantee exact simulation of a 49-qubit circuit.

SA uses a geometric cooling schedule with 1,000 sweeps per read and 1,500 reads per partition per round, totaling $1.5\times10^6$ sweeps per partition. In preliminary tuning, increasing the sweep count beyond 1,000 yielded diminishing energy improvements due to thermalization saturation, while generating 1,500 independent Markov chain reads provided comparable serial CPU runtime to the acquisition window of 25,000 quantum measurement shots. Because quantum projective measurements sample independent configurations without sequential equilibration sweeps, the shot count reflects hardware execution throughput rather than matched algorithmic operations. Candidate ranking and the outer stopping criteria are shared across sampling methods.

\begin{figure*}
\centering
\begin{minipage}[t]{0.48\textwidth}
  \centering
  \includegraphics[width=0.8\linewidth]{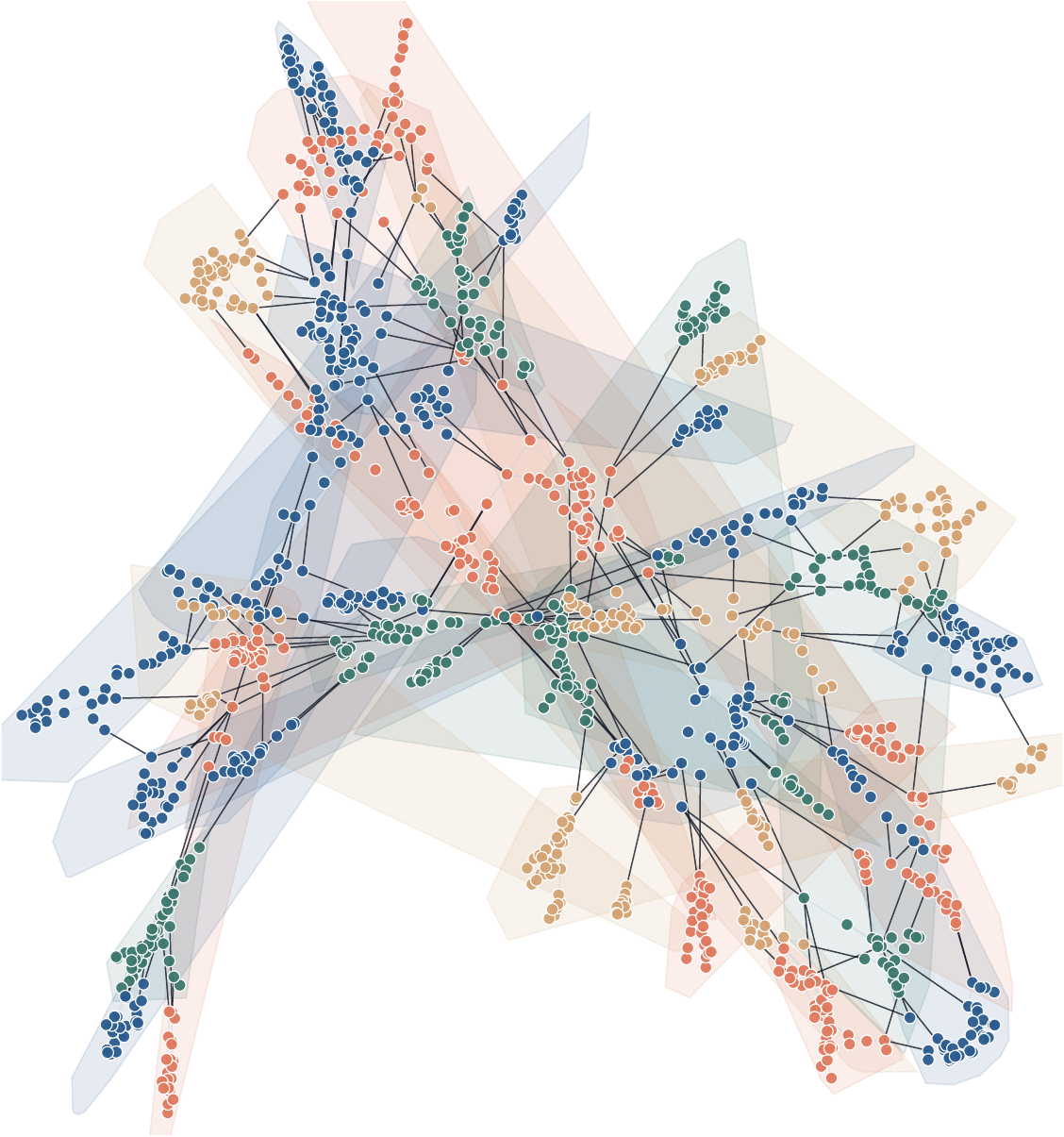}
  \caption{PEGASE-1354 partitioned into 41 subgraphs with edge bound $p_s=50$. Colors distinguish neighboring partitions; gray lines are internal edges and dashed red lines are cut edges. Shaded regions illustrate individual partitions.}
  \label{fig:pegase1354_partitions}
\end{minipage}\hfill
\begin{minipage}[t]{0.48\textwidth}
  \centering
  \includegraphics[width=0.65\linewidth]{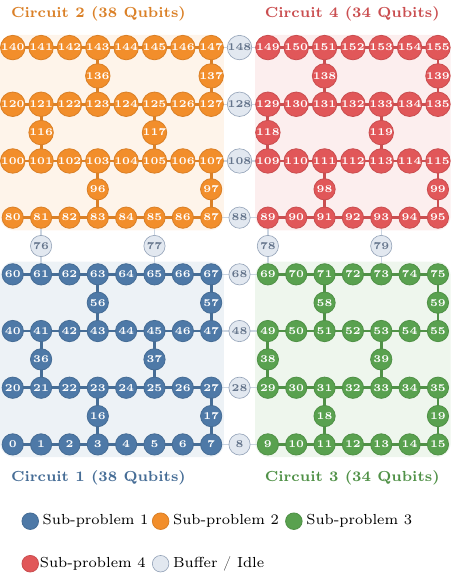}
  \caption{Spatial packing on the 156-qubit IBM Heron processor \texttt{ibm\_marrakesh}. Two 38-qubit and two 34-qubit circuits occupy 144 qubits in disjoint regions. Idle qubits and unused cross-boundary couplers separate the circuits.}
  \label{fig:heron_circuit_packing}
\end{minipage}
\end{figure*}

During hardware benchmarking on \texttt{ibm\_kingston} and \texttt{ibm\_marrakesh}, we evaluated dynamical decoupling \cite{viola1999dynamical} with an $X_pX_m$ pulse sequence and Pauli gate twirling with 32 randomizations \cite{ezzell2023dynamical}. However, for shallow $p=1$ QAOA sampling across distributed partitions, the additional pulse overhead and circuit duration increased effective decoherence and sampling variance without improving the downstream repaired candidate quality. We therefore used Qiskit's heuristic transpilation passes (optimization level 3) focused strictly on minimizing SWAP routing and two-qubit depth, leaving feasibility enforcement to the classical greedy repair routine.

Writing $c_k=\|D_k-D^*\|_F^2/n^2$, the loop stops when $c_{k+1}\le10^{-6}$, when $|c_{k+1}-c_k|/c_k\le10^{-3}$ for $c_k>0$, or after 20 iterations. Decomposition length is $|\{l:\alpha_l>0\}|$. Thus a run's round count and its final horizontal coordinate in the convergence plots need not coincide.

\begin{algorithm}
\small
\SetAlCapSkip{3pt}
\caption{Partitioned E-FCFW Pipeline}
\label{alg:partitioned_qaoa}
\SetAlgoLined
\KwIn{Demand matrix $D^*$, convergence tolerance $\epsilon$, max iterations $K$, sample count $d$, partition bound $p_s$.}
\KwOut{Sparse decomposition $\{(\alpha_l, M_l)\}_{l=1}^{|\mathbb{M}|}$.}
Initialize $\mathbb{M} \leftarrow \emptyset$, $D_0 \leftarrow 0$\;
\For{$k = 0, 1, \dots, K-1$}{
    Construct residual graph $G_k$ from $D^* - D_k$\;
    Partition $G_k$ into subgraphs $\{G_1, \dots, G_m\}$ with at most $p_s$ edges using recursive spectral partitioning\;
    \For{\textbf{each} subgraph $G_i$ \textbf{in parallel}}{
        Formulate QUBO Hamiltonian $\hat{H}_{C,i}$\;
        Execute shallow QAOA circuit with pre-selected parameters $(\gamma, \beta) = (\gamma_0, \beta_0)$; sample candidate set $S_i$\;
    }
    Recombine local samples via beam search or Cartesian stitching\;
    Apply greedy repair; retain the top $d$ distinct matchings as $\mathbb{M}^{\text{sample}}$\;    
    Compute a global maximum-weight matching $M^{MW}$ on $G_k$\;
    Update $\mathbb{M} \leftarrow \mathbb{M} \cup \{M^{MW}\} \cup \mathbb{M}^{\text{sample}}$\;
    Solve cardinality-constrained QP over $\mathbb{M}$ to update weights $\alpha$ and iterate $D_{k+1}$\;
    Prune matchings with zero weight ($\alpha_l = 0$) from active set $\mathbb{M}$\;
    \If{$\|D_{k+1} - D^*\|_F^2 / n^2 \le \epsilon$ \textbf{or} cost improvement has plateaued}{
        \textbf{break}\;
    }
}
\end{algorithm}

\subsection{Stitching and Feasibility Repair}

Cartesian stitching evaluates cross-products in local batches capped at $5^5$ combinations per step. The alternative beam search groups subgraphs hierarchically, retains at most 200 candidates per group, and prunes pairwise combinations back to this limit at each merge-tree level. Both methods limit the otherwise exponential number of combinations.

Raw stitched candidates frequently violate matching feasibility due to two distinct sources of collisions: (i) intra-partition collisions, where shallow circuits and finite QUBO penalties ($\lambda=0.2$) permit adjacent edges to be sampled within a partition; and (ii) boundary collisions, where independent subproblems select edges incident on the same shared cut vertex across partition boundaries.

Feasibility repair resolves these conflicts via a greedy pruning pass over the candidate edge set $\tilde E = \{e_j \in E : x_j = 1\}$ extracted from the stitched bitstring:
\begin{enumerate}
    \item The edges in $\tilde E$ are sorted in descending order of current residual weights $w(e_j) = (D^* - D_k)_{uv}$, with ties broken deterministically by edge index.
    \item An initially empty matching $M \leftarrow \emptyset$ and an occupied vertex set $V_{\text{occ}} \leftarrow \emptyset$ are initialized.
    \item Each edge $e = (u,v) \in \tilde E$ is evaluated in sorted order: if $u \notin V_{\text{occ}}$ and $v \notin V_{\text{occ}}$, the edge is accepted ($M \leftarrow M \cup \{e\}$) and its endpoints are marked ($V_{\text{occ}} \leftarrow V_{\text{occ}} \cup \{u,v\}$); otherwise, it is discarded.
\end{enumerate}
Unselected edges ($x_j=0$) are never introduced, ensuring repair acts strictly as a conflict-resolution filter on the quantum candidate rather than an augmenting heuristic. The procedure runs in $O(|\tilde E|\log|\tilde E|)$ time and guarantees matching feasibility. While repair alters the raw sampling distribution, candidate quality is strictly bounded by the edge set $\tilde E$ proposed by the quantum sampler, which is why structured QAOA samples outperform uniform random samples under the identical repair procedure.

\subsection{MPS Mirror Diagnostic}
\label{app:mirror}
We assess MPS truncation error using Maestro's native mirror fidelity check (\texttt{maestro.mirror\_fidelity}), which implements a deterministic compute-uncompute diagnostic conceptually related to the mirror-circuit principle \cite{proctor2022measuring}.

For a QAOA state-preparation circuit $U$ on $N = |E| = 132$ qubits (comprising initial Hadamard gates applied to $|0\rangle^{\otimes N}$ followed by the parameterized cost and mixer layers), Maestro appends the adjoint circuit $U^\dagger$ in reverse order. In exact execution, $U^\dagger U |0\rangle^{\otimes N} = |0\rangle^{\otimes N}$, returning the initial state with unit probability. Under MPS simulation at bond dimension $\chi$, singular-value truncations during both forward and reverse passes introduce approximation errors that reduce the return probability.
Because full statevector inner products are intractable on 132 qubits, Maestro evaluates this diagnostic empirically via shot-based sampling in the computational basis. Over $S = 15{,}000$ measurement shots, the mirror fidelity is the observed frequency of the all-zero bitstring:
\begin{equation}
\hat{\mathcal F}_{\mathrm{mirror}}(\chi) = \frac{N_{0^{\otimes N}}}{S},
\end{equation}
where $N_{0^{\otimes N}}$ is the count of shots yielding $0^{\otimes N}$. Departures from unity diagnose entanglement truncation in the MPS representation. As noted in Section~\ref{subsec:exp_mps}, this return probability characterizes the accuracy of the MPS approximation on the compute-uncompute trajectory rather than certifying the forward state directly, and is analyzed alongside downstream optimization cost.

\subsection{Orchestration and Circuit Scheduling}

Divi represents residual subgraphs as \texttt{MaxWeight}\allowbreak\texttt{MatchingProblem} instances within a \texttt{Partitioning}\allowbreak\texttt{ProgramEnsemble}, dispatches them through \texttt{QoroService}, and coordinates reconstruction and repair. Qoro's scheduler orders circuits by size and packs them into execution rounds with a shared shot count. It can replicate circuits with large shot requirements on remaining qubits and adjust shot allocation to avoid sparsely filled final rounds. Disjoint register regions and idle buffer qubits separate packed circuits. This reduces dispatch rounds by up to $4\times$ in the reported configurations; the observed PEGASE execution-time reduction is approximately $3\times$.

\subsection{Partition and Packing Layouts}
\label{app:deployment}

Figures~\ref{fig:pegase1354_partitions} and \ref{fig:heron_circuit_packing} expand the partitioning and packing stages of the workflow illustrated in Fig.~\ref{fig:partitioning_workflow}. The PEGASE-1354 layout shows 41 subgraphs, with a four-coloring of the quotient graph to distinguish the subgraphs, emphasizing the splitting up of the workload at the level of the principal problem. The packing layout places four independent circuits (corresponding to different subgraphs) on one Heron processor to be run simultaneously within the same QPU run for a higher ratio of QPU utilization. Spatial circuit packing was possible in this instance because the samples of one subgraph do not sequentially depend on those of another.

\end{document}